\documentclass[a4paper, reprint, aip, rsi]{revtex4-1}
\usepackage{amsmath}   
\usepackage{graphicx}
\usepackage{hyperref}
\usepackage[utf8]{inputenc}
\usepackage[english]{babel}

\begin{document}

\title{Four-contact impedance spectroscopy of conductive liquid samples}

\author{Damjan Pelc}
\email{damjan.pelc@gmail.com}
\author{Sanjin Marion}
\email{sanjin.marion@gmail.com}
\author{Mario Basleti\'c}
\email{basletic@phy.hr}
\affiliation{Department of Physics, Faculty of Science, University of Zagreb, Bijenicka 32, 10000 Zagreb, Croatia}

\begin{abstract}
We present an improved approach to the impedance spectroscopy of conductive liquid samples using four-electrode measurements. Our method enables impedance measurements of conductive liquids down to the sub-Hertz frequencies, avoiding the electrode polarization effects that usually cripple standard impedance analysers. We have successfully tested our apparatus with aqueous solutions of potassium chloride and gelatin. The first substance has shown flat spectra from $\sim$100 kHz down to sub-Hz range, while the results on gelatin clearly show the existence of two distinct low frequency conductive relaxations.
\end{abstract}

\date{\today}
\maketitle

\section{introduction}

Impedance spectroscopy is an experimental technique which enables insight into a wide range of physical phenomena encompassing several length and time scales specific to different systems.\cite{BDS} Here we discuss the low frequency range, for which several types of commercial impedance analysers are available, typically covering ranges from several Hz to several MHz using a self-balancing bridge method. Their basic experimental setup is based on two electrodes that are used both as current and sensing contacts thus enabling a simple geometry and easy sample placement and preparation. Such a setup is then readily modelled with a parallel plate capacitor filled with a conductive dielectric.

\begin{figure}[h!]
\centering
\includegraphics*[width=75mm]{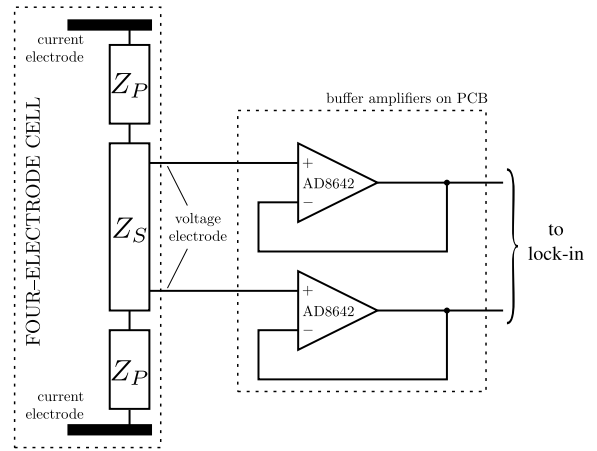}
\caption{Equivalent circuit for the four-electrode cell. $Z_S$ is the sample impedance and $Z_P$ the electrode polarization impedance (one per current electrode). With a sufficient input impedance there is almost no current flowing trough the buffer amplifiers and one is able to ignore the electrode polarizations $Z_P$.}
\label{equiv-circuit}
\end{figure}

Measuring the low frequency impedance response of conductive (and especially
ionic) materials is typically plagued by many problems, the
most important being electrode polarization.\cite{ludijapaneri, bates, sanabria, italijani, mazzeo, HollingsworthSaville} This effect makes the measurement of ionic solution conductivity in a two-electrode
cell practically impossible at frequencies below a few kHz, and several
techniques have been developed to amend this problem. They include
signal analysis for removal the electrode polarization signal from
a two-electrode spectrum,\cite{oduzimanje_pozadine, bates, italijani, HollingsworthSaville} variation of the interelectrode distance
\cite{promjena_udaljenosti, HollingsworthSaville} and four contact configurations.\cite{ludijapaneri, schwan, komparator_bernengo, MayersSaville, mazzeo} However, the only reliable
method for obtaining spectra in the Hertz frequency range seems to
be a four-contact measurement, first proposed by Schwan \textit{et al}.\cite{schwan} It is known that electrode polarization occurs mainly on the electrodes that pass current trough the sample; by using different electrodes for voltage sensing, electrode polarization effects can be avoided (Fig.\ \ref{equiv-circuit}). Although it is possible to extract true bulk low-frequency spectra using model equivalent circuits for the electrode polarization, this is tenuous and often leads to unsatisfactory and/or questionable results.\cite{macdonald} Usually, the impedance relaxation is orders of magnitude smaller than the electrode polarization in the Hz frequency range, making four-contact measurements necessary to obtain credible results due to the limited sensitivity of the experimental setup. No amount of signal analysis can reproduce the quality of a clean spectrum without electrode effects.

There have been several newer designs of four-contact impedance cells (although very few recently), but while yielding good spectra
at low frequencies, they all suffer from large cell volumes,
complicated electronics, impractical sample exchange systems and clumsy
temperature control. 

In this paper we describe a novel planar four-contact
cell design which avoids most of these problems, making accurate measurements
with very small volumes and standard electronic instrumentation possible.
Using a modern current source, lock-in amplifier and high impedance
buffers, our cell is capable of measuring a very wide range of conductivities in the low frequency range,
bringing new physics -- in materials as diverse as gels, biopolymer solutions
and liquid crystals -- within reach.

\section{Apparatus design}

\begin{figure}[h!]
\centering
\includegraphics*[width=70mm]{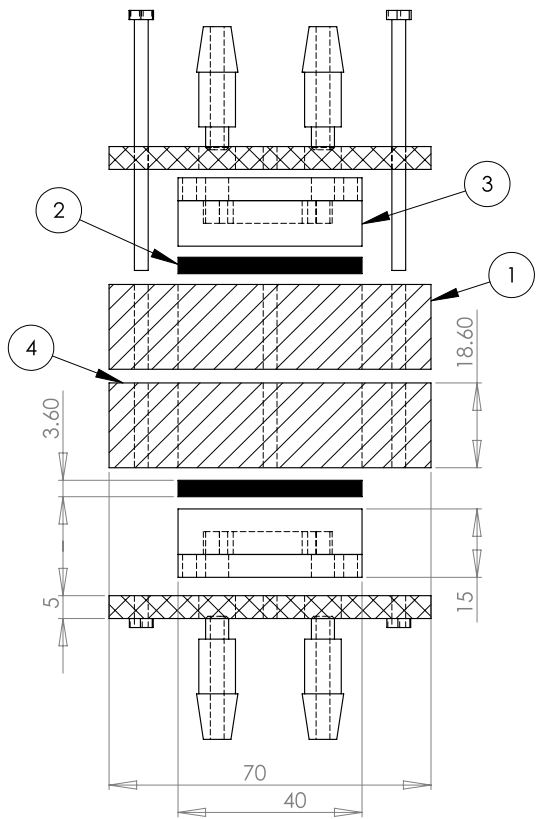}
\caption{Schematic diagram of our impedance spectroscopy cell: 1-polyacetal housing, 2-Peltier element, 3-heat exchange blocks, 4-cell compartment. Dimensions are in millimetres.}
\label{the_cell}
\end{figure}

Our impedance cell consists of two parts: the circuit board used for
impedance measurements and a temperature regulation system based on
two Peltier elements. The layout of the temperature regulation is
shown in Fig.\ \ref{the_cell}, the essential parts being the two Peltier elements,
two aluminium heat exchanger blocks, and a polyacetal housing. A dedicated
heat exchange system was constructed using a radiator for computer
water cooling with a mounted fan and a pump connected to
the aluminium blocks with silicone tubing. The heat exchange liquid
(a 1:1 mixture of water and ethylene glycol, with added anticorrosives)
circulates through S--shaped channels inside the aluminium blocks
at a flow rate of about 5 L/min, yielding enough heat transfer for
the Peltier elements to provide a full 50$^\circ$C difference
between hot and cold sides at maximum rated current. Aluminium blocks
were connected to a common ground to provide shielding for the cell.
The temperature was measured with a platinum thermometer mounted directly
on the impedance printed circuit board (PCB), about 2 mm from the sample itself. A 6.5--digit
Keithley multimeter was used to measure the resistance of the PT-1000
platinum resistor, providing relative temperature sensitivity of at least
1 mK. The temperature was regulated using a variable current source
to power the Peltier elements driven from a PC using a PID controller
written in LabView; temperature stability was typically better than
$0.02$ K, with minimum and maximum working temperatures of $-20^\circ$C
and $80^\circ$C, respectively. The low thermal inertia
of the system and fast current adjusting times enable temperature
quenches with rates up to $150$ K/min, making possible the investigation
of nonequilibrium materials. The precision of our temperature regulation
and the versatility of the system makes it ideal for detailed
and precise temperature-dependent measurements, which are essential
for the study of various transition phenomena. 

\begin{figure}[h!]
\centering
\includegraphics*[width=80mm]{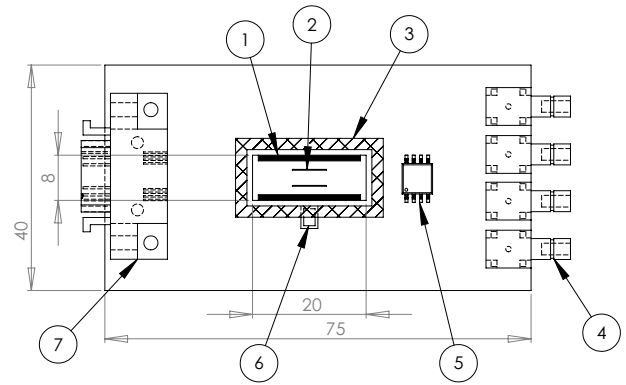}
\caption{Printed circuit board (PCB) schematic of the sensing element for the impedance spectroscopy cell (inserted into region 4 in Fig.\ \ref{the_cell}). 1-gold plated current electrodes, 2-gold plated voltage electrodes, 3-polycarbonate spacer, 4-SMB connectors for the four-contact measurement, 5-buffer amplifier, 6-platinum sensor, 7-board power supply and temperature sensor connector. Dimensions are in millimetres.}
\label{shematic-pcb}
\end{figure}

The heart of our apparatus is the dedicated printed circuit board (PCB) for impedance measuring, depicted in Fig.\ \ref{shematic-pcb}. The current and voltage electrodes are in a planar
configuration which ensures high geometry precision
while making measurements with small sample volumes possible. The
volume of our sample compartment was only 55 $\mu$L, more than an
order of magnitude smaller, compared to configurations in previous works. Such a
small volume guarantees a homogeneous temperature distribution inside
the sample and makes reliable quenching experiments possible. Due to the planar geometry, sample placement and
cleaning is greatly simplified, liquid samples simply being put on
the electrodes with a micropipette. For water solutions, the sample compartment is efficiently cleaned by rinsing with 96\% pure alcohol and deionized water. The circuit board itself was made
from a glass--epoxy composite material to minimize water absorption and covered
with a high resistance lacquer coating wherever possible. Both current (1 in Fig.\ \ref{shematic-pcb})
and voltage electrodes (2 in Fig.\ \ref{shematic-pcb}) are gold plated to avoid electrolysis and
degradation. The board is in direct contact with the lower Peltier
element to ensure good heat transfer, and the thickness of the board
(1 mm) is as small as possible in order to minimize heat resistance
between sample (and thermometer) and Peltier element. To ensure geometry
repeatability and to eliminate artefacts at low frequencies, channels
were machined into the board to hold the sample in place, and a second
piece of the board material was glued to the upper Peltier element,
serving as a lid for the sample compartment. To make the cell airtight
and eliminate evaporation problems an additional polycarbonate spacer (3 in Fig.\ \ref{shematic-pcb})
was machined and glued to the board. 

A crucial detail in making the four-contact cell immune to electrode
polarization is ensuring that no current flows through the voltage
electrodes. Typical lock-in amplifier inputs however have impedances
up to 100 M$\Omega$, making the use of high impedance buffers a necessity.
Our buffer was a two-channel JFET operational amplifier (Fig.\ \ref{equiv-circuit}), AD8642, in
a surface mount package, positioned as close to the voltage electrodes
as possible (4 in Fig.\ \ref{shematic-pcb}) and provided with guard rings for the input pins to eliminate
current leakage. The buffer input impedance of the order of a T$\Omega$
completely eliminates electrode effects allowing, in addition, the measurements of very high sample resistances.
Although at elevated working
temperatures the input bias current of the buffer increases, over
our temperature range it is smaller than $\sim1$ pA,\cite{ad8642} giving a negligible electrode
polarization even at 70$^\circ$C. The buffer amplifier
does not perform voltage subtraction (each voltage electrode had its
own independent buffer channel); this was done with the lock-in differential
input due to a very high specified CMRR. A SR830 digital lock-in amplifier
from Stanford Instruments was used to measure voltages at the voltage
electrodes, while the current was supplied with a Keithley 6221 AC
current source. Such a setup eliminated the need for additional current
measuring circuitry and increased cell simplicity, while also providing straightforward and precise sample impedance determination. With
this instrumentation we were able to measure impedance through six
decades in frequency ($0.1$ Hz to $100$ kHz, the limits being due to prohibitively
long filtering times at the lower end).

\begin{figure}[h!]
\centering
\includegraphics*[width=80mm]{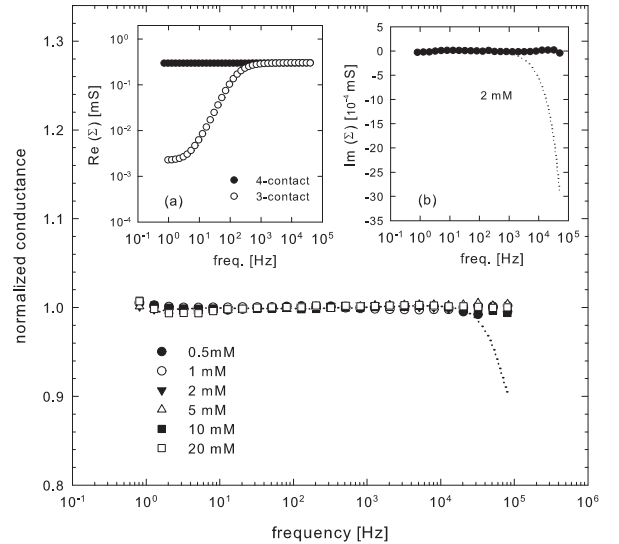}
\caption{Normalized conductance of different concentrations of potassium chloride aqueous solutions. Sample volume was $55$ $\mu$L and all spectra were normalized to their average value. Dashed lines show a sample of the raw data without taking account of the inadequacies of the current source, lock-in amplifier and the length of the cables in the system.  Inset (a) shows the comparison with a 3-contact measurement which does not exclude polarisation effects on one of the current electrodes.  Inset (b) is the imaginary part of the conductance.}
\label{calibration_graph}
\end{figure}

\section{Assessment and discussion}

Before any test measurements were made we identified the characteristic of our setup without the sample chamber (raw data shown as a dotted line in Fig.\ \ref{calibration_graph}) as is common procedure in impedance spectroscopy. The raw spectrum is flat up to $\sim 30$ kHz, but for higher frequencies deviations are observed. Careful analysis revealed that they are a consequence of deficiencies in the synchronization signal between instruments (Keithley 6221 current source and SR830 lock-in) at higher frequencies, and we found that this is highly reproducible for this particular choice of instruments. Accordingly, all subsequent measurements results were adjusted to account for these instrument--induced deviations.

The cell was tested and calibrated using standard potassium chloride water solutions in a wide
conductivity range. On the calibration graphs for the conductivity (Fig.
\ref{calibration_graph}) it can be clearly seen that there are almost no spurious effects
in the spectra; their flatness is typically better than 1\% over five
decades in frequency. This is in stark contrast to two-electrode measurements,
and for comparison we include a spectrum obtained by measuring the
voltage on one voltage electrode (`three-electrode' configuration,
Fig.\ \ref{calibration_graph} insert (a)) which is similar to a two-contact spectrum (in a real
two-contact measurement the electrode polarization effects would be
larger for another factor of two). The imaginary part of the conductance was much more difficult to measure accurately due to the fact that it is, in this case, orders of magnitude smaller than its real counterpart. This problem might be reduced by using a lock-in amplifier with an incorporated compensation. A typical imaginary part spectrum is shown in insert (b) of 
Fig.\ \ref{calibration_graph}, including the raw data without any setup compensation (dotted line).

\begin{figure}[h!]
\centering
\includegraphics*[width=80mm]{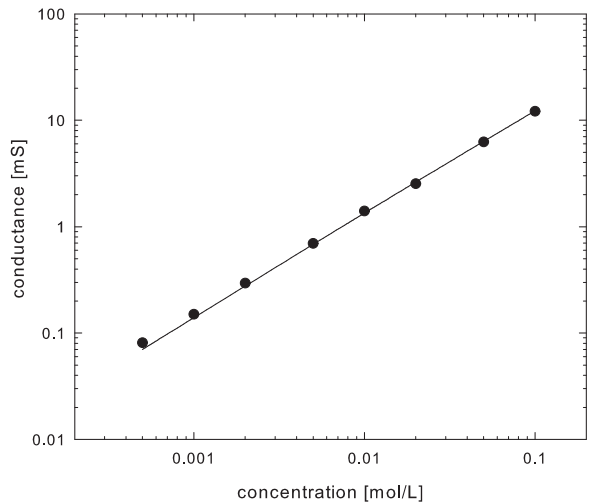}
\caption{Concentration dependence of the conductance of the aqueous potassium chloride solutions. The solid line is obtained from literature values by multiplication with the cell constant, $9.6$ mm.}
\label{concentration_conductivity}
\end{figure}

The plot of potassium chloride solution conductivity against solution
concentration (Fig.\ \ref{concentration_conductivity}) proves that the cell constant does not depend
on the sample resistance, justifying the validity of our planar layout. The cell constant
was calculated to be $(9.6\pm 0.1)$ mm using conductivity values from the literature\cite{CRC} and is in good agreement with the true cell layout, main source of error being the sample volume uncertainty. To minimize possible electrolytic and nonlinear effects at low frequencies
we were careful to always keep the voltages across the sample
smaller than 10 mV, with corresponding currents in the $\mu$A
range. Linearity was confirmed by checking the second and third harmonic
signal levels (which were always below the noise level) and the uniform conductivity
at low frequencies so we can conclude that electrolytic effects were negligible. 

Although providing a simple measurement arrangement, the main drawback of our cell is the aforementioned problem with reduced sensitivity in the imaginary part of the conductivity. A second issue is the sample geometry; if the sample is not fairly
symmetrically placed on the electrodes, artefacts will appear in the
spectra. We tried to minimize this by making the sample compartment
roughly quadratic and providing edges to keep a liquid drop in place
with surface tension. This works out fine for water solutions, but
low surface tension samples tend to spill and ruin the geometry. This
problem was amended by adding a quadratic  polycarbonate liquid holder inside the sample compartment, with a net volume of 73 $\mu$L. Low surface
tension liquids could then be made to fill the polycarbonate compartment
completely. However, for aqueous samples, which are generally
more problematic due to electrode polarization (and thus more sensitive
to geometry) better results were obtained by using channel edges to
hold the liquid in place instead of the polycarbonate walls. 

\begin{figure}[h!]
\centering
\includegraphics*[width=90mm]{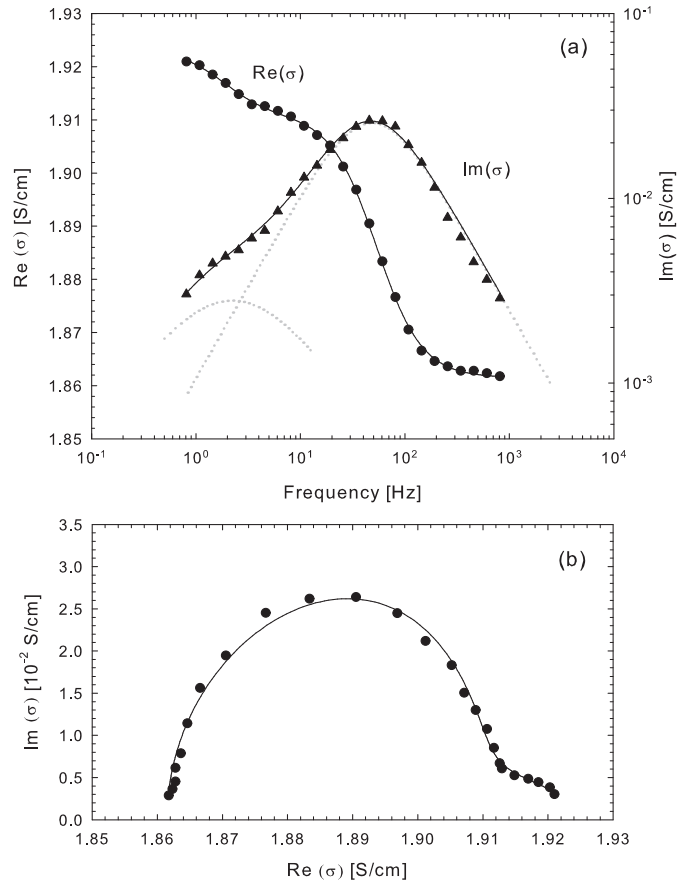}
\caption{(a): Impedance spectra for a $1.5$\% gelatin sample at $13^\circ$ C. Real and imaginary part of the complex conductivity are shown. Solid lines are fits to two Cole-Cole relaxations\cite{colecole}. The two dotted lines show the individual contributions of the two relaxation processes in $\mathrm{Im}(\sigma)$. (b): Complex plane plot of the relaxation in conductivity.}
\label{gelatin-complex}
\end{figure}

To illustrate the possibilities for application of such a cell, we
perform a measurement of the conductivity spectrum of an aqueous solution
of pig skin gelatin (Sigma-Aldrich, type A, Bloom 300). The gelatin solution was prepared by dissolving in deionized water and stirring at 60$^\circ$C for 2 hours.
Fig.\ \ref{gelatin-complex} shows the conductivity spectrum of a $1.5$\% gelatin solution in
water at 13$^\circ$C. In this experiment two low-frequency
conductivity relaxation processes are revealed for the first time,
and they can be clearly resolved in both the real and imaginary parts. The experimental data was fitted to a superposition of two Cole-Cole\cite{colecole} relaxations in the conductivity spectrum yielding an accurate fit.

 The shape of the curves (the real part of the conductivity decreases near the
relaxation frequency, while we would expect it to increase in a dielectric
process) and comparison of our data with other techniques (light scattering\cite{anomalous_diffusion} and diffusion NMR\cite{diffusionNMR}) suggests that we are able to detect the translational
diffusion of the charged gelatin chains; the chain contribution to the
conductivity is turned off at the relaxation frequency because the
field is varying too quickly for them to follow it. We believe that the faster relaxation process corresponds to chain segment fluctuations while the slower process is due to whole chain reptation, in agreement with conclusions from light scattering experiments.\footnote{Detailed account to be published elsewhere.} We emphasize that a direct measurement of the charged chain diffusion is a result of our measurement method which avoids problems due to the polarization effects; the height of the relaxation is in this case only about 1\% of the asymptotic conductivity of the sample and would thus be unattainable using a standard two-contact cell and signal analysis methods (e.g. fitting a superposition of an electrode polarization model and two Cole-Cole relaxations).

\section{Conclusion}

In conclusion, we have presented a small volume four-contact
impedance spectroscopy cell for use primarily with ionic water solutions. The low
frequency range ($1 - 10^{5}$ Hz) opens up possibilities not accessible
to standard two-contact impedance spectroscopy, including the observation
of large scale motions of polyelectrolytes (as we have demonstrated
with a gelatin-water solution) useful for a wide range of biologically
and industrially interesting systems. Besides protein and polypeptide
solutions, notable possible applications are DNA, hyaluronic acid and similar
biological polyelectrolytes, charged polymers like xanthan,
protein aggregations and even larger structures like emulsions or
amphiphylic systems. Due to the simplicity and small size, we believe that our cell
represent a convenient device/tool for investigating such systems with low frequency impedance spectroscopy.

\begin{acknowledgments}
We are pleased to acknowledge the assistance of prof.\ dr.\ sc.\ Miroslav Po\v{z}ek, dr.\ sc.\ Tomislav Vuleti\'c, dr.\ sc.\ Nata\v{s}a \v{S}ijaković-Vuji\v{c}i\'c and prof.\ dr.\ sc.\ Amir Hamzi\'c. This work was partially funded by the Croatian MZOS Projects 119-1191458-1023 and 119-1191458-1022, and UKF project number 55.
\end{acknowledgments}

\end{document}